\documentstyle[aps,epsf]{revtex}  

\def\fs{$f_S$}
\def\Journal#1#2#3#4{{#1} {\bf #2}, #3 (#4)}
\def\NCA{\em Nuovo Cimento}

\def\NIMA{{\em Nucl. Instrum. Methods} A}
\def\NPB{{\em Nucl. Phys.} B}
\def\PLB{{\em Phys. Lett.}  B}
\def\PRL{\em Phys. Rev. Lett.}
\def\PRD{{\em Phys. Rev.} D}
\def\PRO{\em Phys. Rev.}
\def\PREP{\em Phys. Reports}
\def\ZPC{{\em Z. Phys.} C}
\def\sqs{$\sqrt s$}

\begin{document}        

\baselineskip 14pt
\title{Diffractive Results from the Tevatron}
\author{Gilvan A. Alves}
\address{Lafex/CBPF \\ Rua Xavier Sigaud 150, 22290-180, Rio de Janeiro, RJ, Brasil}
\author{For the D\O\ and CDF Collaborations}   
%
\maketitle              

\begin{abstract}        
    Hard diffraction in events with dijets and rapidity gaps has been studied
by D\O\ and CDF for three processes: hard color singlet exchange, hard single diffraction,
and hard double pomeron exchange, using Tevatron 
$\overline  pp$ data at $\sqrt{s}$ = 
 630 GeV and 1.8 TeV. Measurements of rates,
$\eta$, $E_T$ and $\sqrt{s}$ dependencies are presented and comparisons made 
with predictions of several models.

\
\end{abstract}   	

\section{Introduction}               

Events with a region of rapidity space devoid of particles (rapidity gaps)
were first observed in cosmic ray data \cite{Cosmic}.  
The idea of diffractive
dissociation of projectile and target to produce such events 
soon followed \cite{Good}.
Later the interpretation of total, elastic and diffractive cross sections
in terms of the exchange of an object with the quantum numbers of the vacuum,
called the pomeron, proved very useful \cite{Dino}.
F. Low and S. Nussinov suggested that the pomeron corresponds to
the interchange of two gluons \cite{Low}. In a latter development 
Ingelman and Schlein (IS) proposed that high $p_T$ jets could be diffractively
produced via pomeron exchange and that this might probe
the partonic structure of the pomeron \cite{Ingelman}.
Events containing rapidity gaps and jets were
first observed by UA8 \cite{UA8}, giving rise to the field of hard diffraction. 
This area of interest has expanded considerably in the last decade, with
the availability  of high energy colliding beams. 
Jet production with rapidity gaps have been observed 
at the Tevatron \cite{Tevatron,CDF2} and at HERA \cite{HERA} .
CDF and D\O\ have  studied dijet + rapidity gap events using data 
from the 1992-1996 Tevatron
$\overline  pp$ collider run (Run I) at center-of-mass energy (\sqs~) = 
 1.8 TeV, with a short run at 630 GeV during that period.
 Because of the large
center of mass energy and large integrated luminosity,
the new CDF and D\O\ results can give further insight into 
diffractive processes.

 The D\O\ detector is described elsewhere \cite{D0_det}. 
 Jets are found in the uranium-liquid argon calorimeters
 using a cone algorithm with radius $R = 0.7$ 
 in the $\eta-\phi $ plane \cite{geom}.
 Particle multiplicity is determined in the central region
($|\eta| < 1.0$) using the number of towers 
 ($0.1 \times 0.1$ in $\Delta\eta \times \Delta\phi$)
with transverse energy ($E_T$) above 200 MeV in the central 
electromagnetic calorimeter and the number of tracks in the 
central drift chamber. In the forward region this multiplicity
is measured by the number of towers with ($E_T$) above 125 MeV 
in the electromagnetic end cap calorimeter (2.1 $<|\eta| < 4.1$) 
and 500 MeV in the hadronic end cap calorimeter (3.2 $<|\eta| < 5.2$).
Because the last layer of the hadronic calorimeter (at the limit of 
the forward acceptance) is composed of stainless steel and produces
less noise than the uranium sections, the threshold for particle detection 
was reduced to 50 MeV in this layer.
In addition, in the forward region we also use an array of scintillator
hodoscopes, called L\O\ detector, to tag the presence of charged particles
in the region 2.3 $<|\eta| < 4.3$.

The CDF detector, described in ref. \cite{CDF_det},
consists of a large central detector with
tracking in a solenoidal field and calorimetry over
$|\eta| < 4.2$.  
To measure particle multiplicities,  CDF uses
 the  central tracker ($|\eta| < 1.1,
 ~p_T{^{track}} > 300$ MeV), the central calorimeter
 ($|\eta| < 1.1, ~E_T{^{tower}} > 300$ MeV corrected),
 and forward calorimeters (2.2 $<|\eta| < 4.2$).
 For the last two
 months of the collider run, CDF  installed  three Roman Pot
detectors to trigger on quasi-elastically scatted antiprotons.
Nearly all the pot triggers
have $0.05< \xi < 0.1$, where $\xi$ = $1 - x_{F}$ is  
the fraction of momentum lost by the antiproton and carried by the pomeron.

\section{Hard  Color Singlet Exchange}

Two jets separated by a rapidity gap has been proposed as the signature
of color singlet exchange (CSE) carrying a 
high $Q{^2}$ \cite{Dok,Bj}. 
 Rapidity gaps between jets have
been observed both at the Tevatron \cite{Tevatron} and 
at the DESY $ep$ Collider (HERA) \cite{HERA}.
The measured rates of $\approx 1\% $ at the Tevatron and $\approx 10\% $ at HERA
are too large to be accounted for by electroweak boson exchange and
indicate a strong interaction process.

D\O\ and CDF have made recent studies of dijet data with central 
rapidity gaps. 
Both experiments measure the color singlet fraction (\fs) at
\sqs~ of 630 GeV and 1.8 TeV.
The observed color singlet fraction includes the probability that
the rapidity gap is not contaminated by particles from spectator 
interactions. This survival probability ($S \sim 10\%$ at 1.8 TeV)
is assumed to be
independent of Bjorken $x$ and the flavor of the initial partons in the hard
scattering \cite{F-S,G-L-M_1} but depends on \sqs~
$(S_{630}/S_{1800}=2.2\pm0.2)$ \cite{G-L-M_2}.

CDF measures the fraction of colorless exchange to all opposite side dijets
from the tracking distribution. Results are listed in 
Table~\ref{table:cse_fs}.
The ratio of the CDF fractions from the measurements at the two center of mass
energies is $ R(\frac{630}{1800})   = 2.4\pm 0.9$. No $E_T$ dependence 
of the signal is observed.

\begin{figure}[hbt]
\vspace*{-0.35in}
\centerline{\epsfxsize=3.5in \epsffile{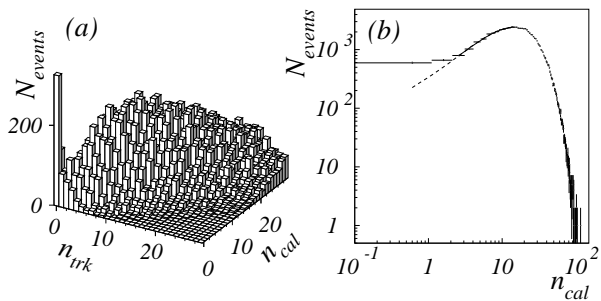}}
\vspace*{-1.5in}
\caption{ The multiplicity between the dijets for the D\O\ high-$E_T$ 
1800~GeV sample: (a) 
  two-dimensional multiplicity, $n_{cal}$ vs. $n_{trk}$; (b)
$n_{cal}$ only with NBD fit, 
plotted on a log-log scale to emphasize low multiplicity bins.
\label{f:1800}
}
\end{figure}
\begin {table}
\begin{tabular}{clclcc}
experiment& $\sqrt{s}$ (GeV) & jet $ E_T$ & $\eta^{jet}$ & nb. of triggered events\\ \tableline
CDF   & ~630 & $>$~8 GeV & 1.8$<|\eta|<$3.5 & ~1k + 1k(same-side) \\
D\O\  & ~630 & $>$12 GeV & $|\eta|>$1.9     & ~7k \\
CDF   & 1800 & $>$20 GeV & 1.8$<|\eta|<$3.5 & 10k +30k(same-side) \\
D\O\  & 1800 & $>$12 GeV & $|\eta|>$1.9     & 48k \\
D\O\  & 1800 & $>$25 GeV & $|\eta|>$1.9     & 21k \\
D\O\  & 1800 & $>$30 GeV & $|\eta|>$1.9     & 72k \\

\end{tabular}
\caption{Kinematic cuts for color singlet exchange.} 
\label{table:cse_kin}
\end{table}

\begin{table}
\begin{center}
\begin{tabular}{lclclcc}   
~~\fs (\% ) & $ E_T $ (GeV) & \sqs (GeV) & experiment \\ \hline
$2.7~\pm0.7~(\rm stat.)\pm0.6~(\rm syst.)$ & $>8$ & ~630 & CDF   \\ 
$1.85\pm0.09(\rm stat.)\pm0.37(\rm syst.)$ & $>12$& ~630 & D\O\  \\ \hline
$0.54\pm0.06(\rm stat.)\pm0.16(\rm syst.)$ & $>12$& 1800 & D\O\  \\
$0.94\pm0.04(\rm stat.)\pm0.12(\rm syst.)$ & $>30$& 1800 & D\O\  \\ 
$1.13\pm0.12(\rm stat.)\pm0.11(\rm syst.)$ & $>$20& 1800 & CDF\\ 
\end{tabular}
\end{center}
\caption{Color singlet Fractions
.}
\label{table:cse_fs} 
\end{table}

D\O\ has recently published results for dijet + central gap events\cite{PLB}.
(See Table~\ref{table:cse_kin} for kinematic cuts.)
Single interaction events are required. 
The particle multiplicity in the central rapidity region
is approximated by the multiplicity, $n_{cal}$,
of 
transverse energy  above 200 MeV
in the electromagnetic  calorimeter,
and by the track multiplicity in the central tracking chamber,
$n_{trk}$. Figure 1 (a) shows the D\O\ multiplicity distribution for
$n_{cal}$ versus $n_{trk}$. 

To calculate the fraction due to color singlet exchange,
the leading edge of each $n_{cal}$ distribution is
fitted using a single negative binomial distribution (NBD).
The fraction of rapidity gap events (\fs) is calculated from
the excess of events over the fit in the first two bins
($n_{cal}=0$ or 1)
divided by the total number of entries. 
Figure 1 (b) shows the $n_{cal}$  distribution
and the NBD fit for the high $E_T$ sample. 
See Table~\ref{table:cse_fs}  for values of \fs.  
The D\O\ value of the ratio of the rapidity gap fractions at 630 and 1800 GeV is
$ R(\frac{630}{1800})   = 3.4\pm 1.2$.

Measuring the color-singlet fraction as a function of $E_T$, $\eta$ and \sqs~
  probes the nature of the color-singlet exchange and 
  its coupling to quarks and gluons.
If the color-singlet dynamics are similar to single gluon exchange
  except for different coupling factors to quarks and gluons,
  the color-singlet fraction would depend only on parton distribution functions 
  via $x_{F}$.
Thus for a color-singlet that couples more strongly to gluons than quarks,
  the color-singlet fraction would fall as a function of increasing 
  $x$, 
  since the gluon distribution becomes suppressed relative to the 
  quark distribution as $x$ increases. This implies a decreasing 
  color singlet fraction with increasing jet $E_T$ and $\Delta\eta$
  or decreasing \sqs.

\begin{figure}
\vspace{-1cm}
\centerline{\epsfysize=2.5in \epsffile{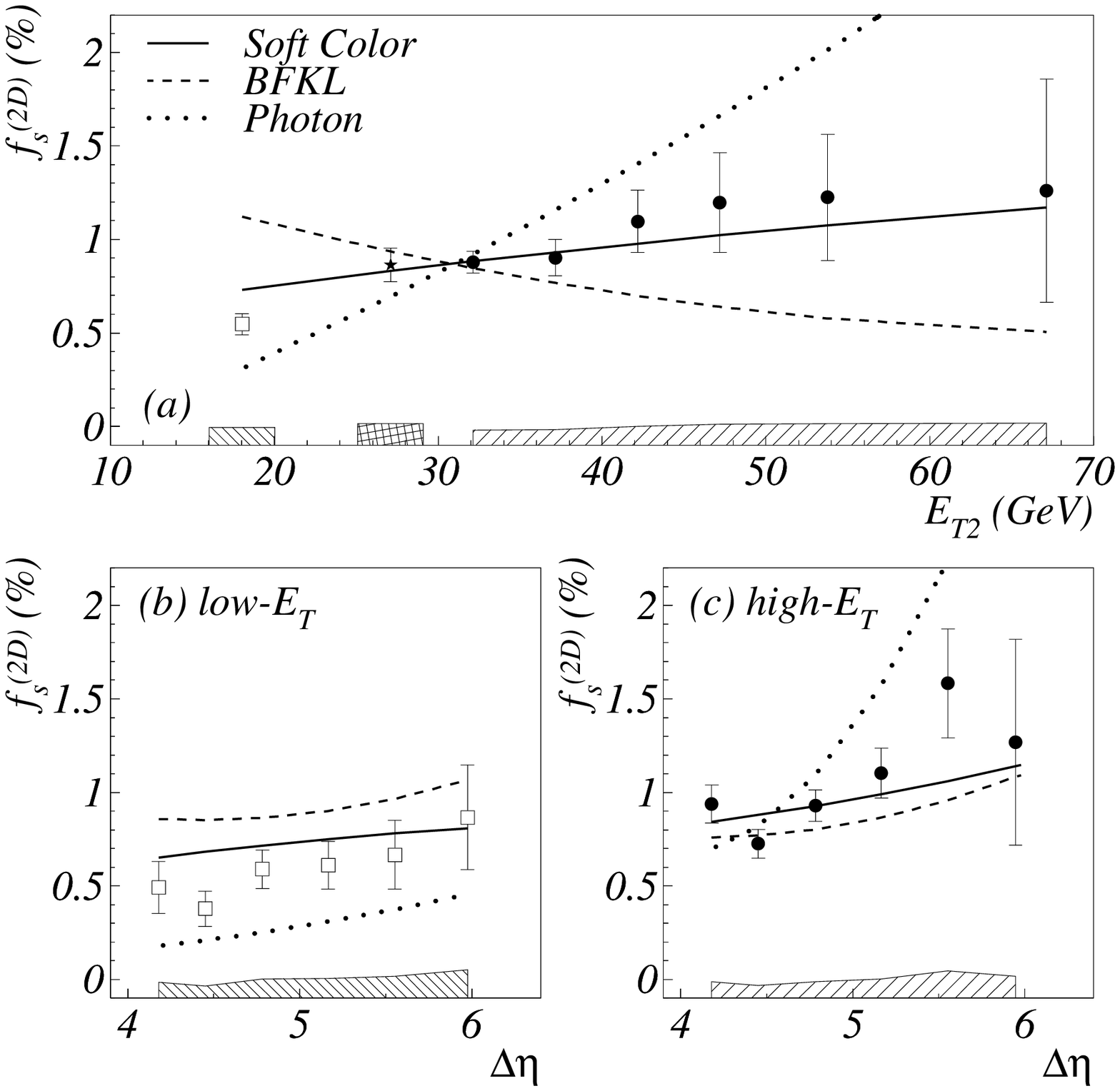}}
\caption{Monte Carlo fits to the measured color-singlet fraction 
         $f_{2D}$.
         The normalization is allowed to float.}
\label{fig:d0_cse_et_}
\end{figure}

   To measure the color-singlet fraction as a function of $E_T$ and $\eta$, D\O\
uses the two-dimensional multiplicity, ($n_{cal}\;{\rm vs.}\;n_{trk}$) 
 which gives improved signal-to-background
ratios  compared to the NBD method. This is useful for smaller statistics 
samples   and avoids large uncertainties in the color-exchange background 
subtraction. 
The ``2D'' color-singlet fraction
$f_{2D}$ is defined as the fraction of events with $n_{cal}+n_{trk}<2$.
The results are shown in Figure~\ref{fig:d0_cse_et_}. 
The systematic errors include effects from background estimation.
The measured color-singlet fraction
shows a slight rise as a function of dijet $E_T$ and $\Delta\eta$.

  To compare the experimental color singlet fractions to models, 
D\O\ uses {\sc HERWIG 5.9} \cite{HW},
which includes a two-gluon exchange with BFKL dynamics \cite{M-T}, and uses CTEQ2M 
parton distribution functions. In addition, D\O\ uses the $t$-channel
photon exchange process in {\sc HERWIG} to investigate  models in which
the color singlet couples only to quarks with a massless photon-like singlet.

In the soft-color rearrangement model\cite{Halzen},
initial state quarks have 
fewer color combinations and thus,
a higher probability of being rearranged
into a colorless state, than initial state gluons, i.e. $C_{gg}<C_{qg}<C_{qq}$,
 where the ``$C_{ab}$s'' are the effective color factors representing the 
couplings to different initial state partons.
A reasonable choice of color factors is $C_{qq}  = \frac{1}{9}$, $C_{qg}  = \frac{1}{24}$ and 
$C_{gg} = ({\frac{1}{64}})$.  
Predictions of these models are simultaneously
fit to the experimental $E_T$ and $\Delta\eta$ dependence of \fs~ at \sqs
= 1.8 TeV, letting the normalization float. The results are shown in Figure~\ref{fig:d0_cse_et_}.
The data favour color-singlet models that couple more strongly to quarks  
than gluons, but a single-gluon model (no dependence) can not be excluded.

\section{Hard Single Diffraction}

In the IS picture of hard single diffraction, a pomeron (color singlet object) 
is emitted
from the incident $p$ ($\overline  p$) and undergoes a hard scattering with
the $\overline  p$ ($p$), leaving a rapidity gap in the direction of the
parent particle.  The signature is   
two jets  produced on the same side and a
forward rapidity gap along the direction of one of initial beam particles.

\begin{figure*}[htb]
\vspace{0.5in}
\vbox{
\centerline{\epsfysize=3.5in \epsffile{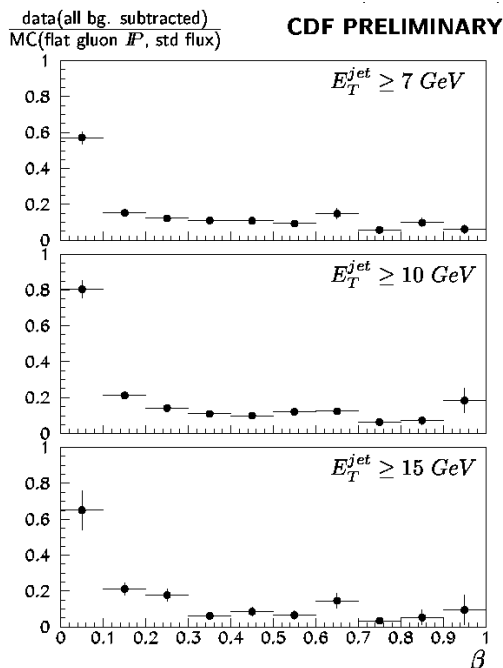}}
\caption{Ratio of Data to Monte Carlo simulation as
a function of $\beta$ using a flat gluon distribution and standard
pomeron flux}
\label{cdf_hsd}
}
\end{figure*}

Data was taken by the CDF detector at the end of Run I using a trigger which requires 
tagging the recoil $\overline  p$ with ``Roman Pot'' detectors. The typical acceptance
for these detectors is $0.05 < \xi < 0.1$ and $0 < |t| < 2~ GeV^{2}$, 
where $\xi$ is the fractional momentum lost by the antiproton and $t$ its four momentum
squared. After applying several cuts to select events with a good reconstructed track 
in the Roman Pots, CDF extracted the momentum fraction of the interacting parton in the
pomeron, $\beta$, for dijet events with $E_{T} > 7 GeV$, using the following expression:

\begin{eqnarray}
{\beta} = {{E^{jet1}_{T} exp(-\eta^{jet1}) + E^{jet2}_{T} exp(-\eta^{jet2})} 
\over {2 \xi P_{beam}}}
\end{eqnarray}

The $\beta$ distribution for the pomeron was obtained by subtracting several
background contributions in the data, of which the most importants are 1) 
non-diffractive dijet events accidentally overlapped with a Roman Pot hit, 2) meson 
exchange background and 3) double diffraction background. After subtracting these 
contributions from the data, then unfolding the detector acceptance by using 
simulations with a flat gluon distribution, the data was divided by Monte Carlo
simulations based on POMPYT\cite{POMPYT} using a flat gluon distribution and the
standard Donnachie and Landshoff flux parametrization\cite{Donn}. The comparison, 
shown in Figure~\ref{cdf_hsd}, shows agreement in shape for $\beta > 0.2$, but 
there is a discrepancy in the normalization by about a factor of 6, as well as an
enhancement for the low $\beta$ region.

\begin{figure*}[htb]
\vspace{-0.5cm}
\centerline{\epsfysize=4.0in \epsffile{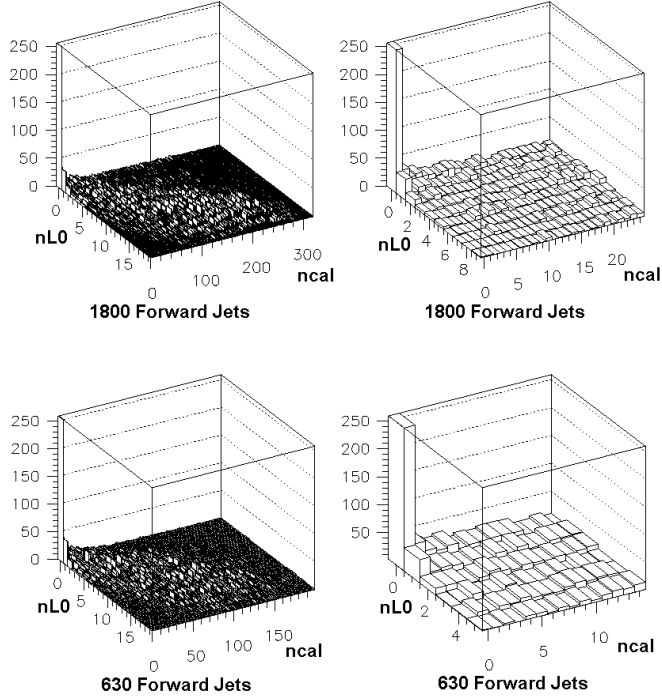}}
\caption{ Number of hits level 0 ($\rm{n_{L0}}$) and calorimeter towers 
($\rm{n_{CAL}}$) 
above threshold opposite to jets in the 
forward trigger sample at center of mass energies $1800$ and $630$\,GeV.}
\label{fig:d0_hds_mult_}
\end{figure*}

The D\O\ data were obtained using an inclusive jet trigger 
or a forward two jet trigger. 
Events are selected with two jets with 
$E_T$ $>$ 12 GeV and  $|\eta| > $1.6. The number of end cap calorimeter towers 
($n_{cal}$) above threshold
is measured opposite to the leading two jets. 
The $n_{cal}$ distribution
for the forward trigger sample at center of mass energies of 1800 and 630 GeV 
is shown
in Figure~\ref{fig:d0_hds_mult_}. A clear peak is seen in the 
$n_{L0}~=~n_{cal}~=0$
(zero multiplicity) bin as expected for a diffractive signal.
A two-dimensional fit on the $n_{L0}~vs.~n_{cal}$ distribution, where data and 
background are fit simultaneously, allows the direct extraction of the fraction 
of events containing a rapidity gap at both energies.
This gap fraction, including statistical and systematic uncertainties, is 
determined to be $0.64 \pm 0.05(stat.~+~syst.) \%$ 
for the 1800 GeV data and $1.23~+~0.10~-~0.09(stat.~+~syst.) \%$ for the 
630 GeV data. Work is in progress for the extraction of the pomeron parton 
distributions.

\section{Hard Double Pomeron Exchange}

Central dijet events containing two rapidity gaps or a rapidity gap on the opposite
   side of a
quasi-elastically scattered anti-proton have been studied by D\O\ and CDF respectively.
This event topology is consistent with double pomeron exchange (DPE).
The data can be used to give more information about the hypothesized pomeron.

\begin{figure*}[htb]
\begin{minipage}{0.45\linewidth}
\begin{center}
\mbox{(a)\epsfxsize 5.5 truecm \epsfbox{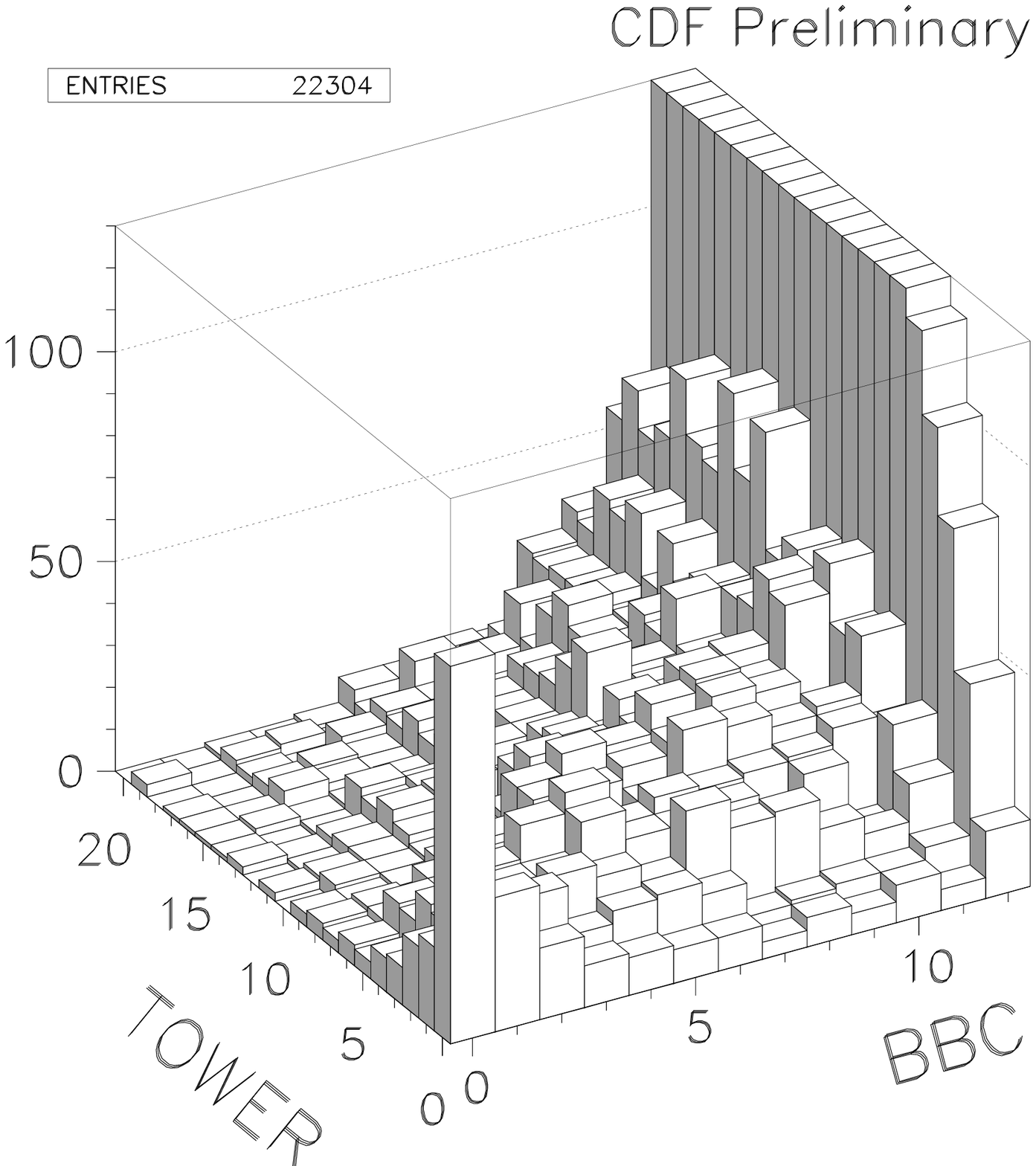}}
\end{center}
\end{minipage}
\hfill
\begin{minipage}{0.45\linewidth}
\begin{center}
\mbox{(b)\epsfxsize 5.5 truecm \epsfbox{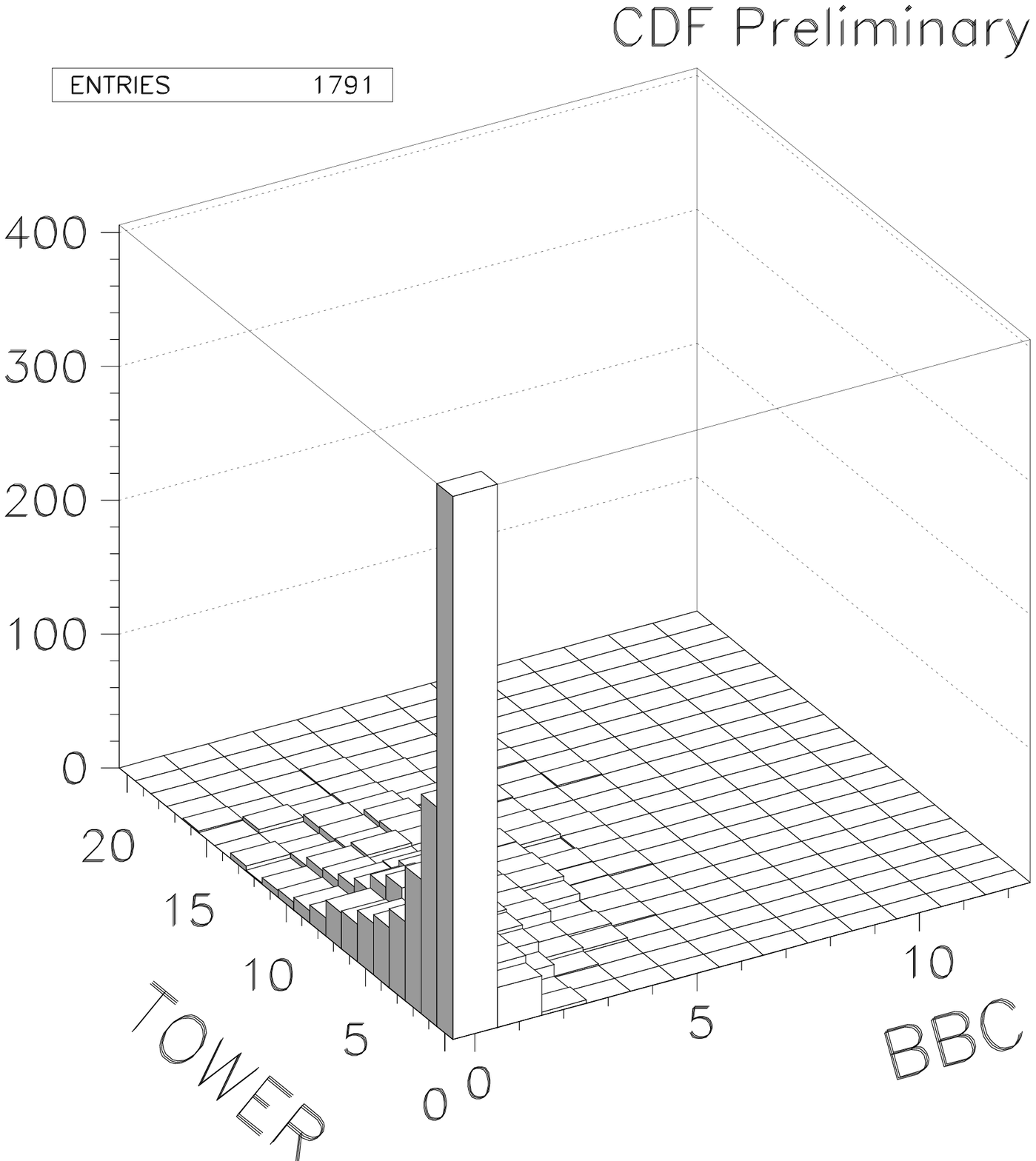}}
\end{center}
\end{minipage}
\caption{(a)
East side BBC vs Cal. tower multiplicity distribution for PJJ events,
and 7 GeV dijets.
(b) East side BBC vs Cal. tower multiplicity distribution for Monte Carlo
Double Pomeron (with a flat $\beta$ distribution.)
} 
\label{fig:cdf_dpe_lego}      
\end{figure*}

   CDF took data at 1800 GeV center of mass energy using the Roman Pot 
trigger to tag antiprotons. 
A sample of 27,000 events with a tagged $\overline  p$ and at least
two jets  with $E_T > 7$ GeV is obtained.
Low multiplicity events are selected by requiring
$N_{BBC}(west)\leq  6$ in the 16 element Beam-beam counters (BBC) 
on the same side as the pots.
This gives 22,304 PJJ (pot-jet-jet) events. A sample of minimum bias events, with the same
dijet selection, is used for comparison.
   Figure~\ref{fig:cdf_dpe_lego} (a) shows the calorimeter tower multiplicity 
in the east side (opposite 
to the pot track), $N_{FCAL}$, versus the number of hits in the BBC.

In order to understand the shape of this distribution, 
Monte Carlo events were generated using a version of POMPYT
modified to include double pomeron exchange, where the incoming  proton
and antiproton emit pomerons with a standard flux (Donnachie and Landshoff form
\cite{Donn}  with parameters as measured by CDF \cite{CDF_94}).
The pomeron-pomeron interaction
is treated like a hadron hadron collision which produces jets.
A flat $\beta$ distribution
of partons inside the pomeron was assumed. Diffractive deep
inelastic scattering data from HERA suggest 
such a hard structure with a rather flat $\beta$-distribution \cite{DDIS}.
 The simulated
events are shown in Figure~\ref{fig:cdf_dpe_lego} (b). The strong signal in 
the (0,0) bin only
contains 24 \% of the DPE events with $\xi <0.1$ for 7 GeV dijets.

By extrapolating linearly into the $N_{BBC} = N_{FCAL} = 0$ bin along the diagonal
axis, CDF obtained the ratio of dijet gap to dijet no-gap events to be:

\begin{center}              
$ R(\frac{PJJG}{PJJ})= [0.36\pm 0.05(stat.)\pm(0.03(syst.)]\%$
\end{center}

\noindent where G means gap (no detected particles for $2.4 <\eta < 5.9$), JJ means two jets
with $E_T > 7$ GeV and P means a pot track with $0.05< \xi < 0.1$. 
    When
the discrepancy factor, D=18\%, found in previous analyses of diffraction 
in $\overline pp$ and $ep$ collisions \cite{CDF2}, is applied (squared),
 the ratio from simulation is in good agreement with the data.

\begin{figure*}[htb]
\begin{minipage}{0.45\linewidth}
\begin{center}
\mbox{(a)\epsfxsize 5.5 truecm \epsfbox{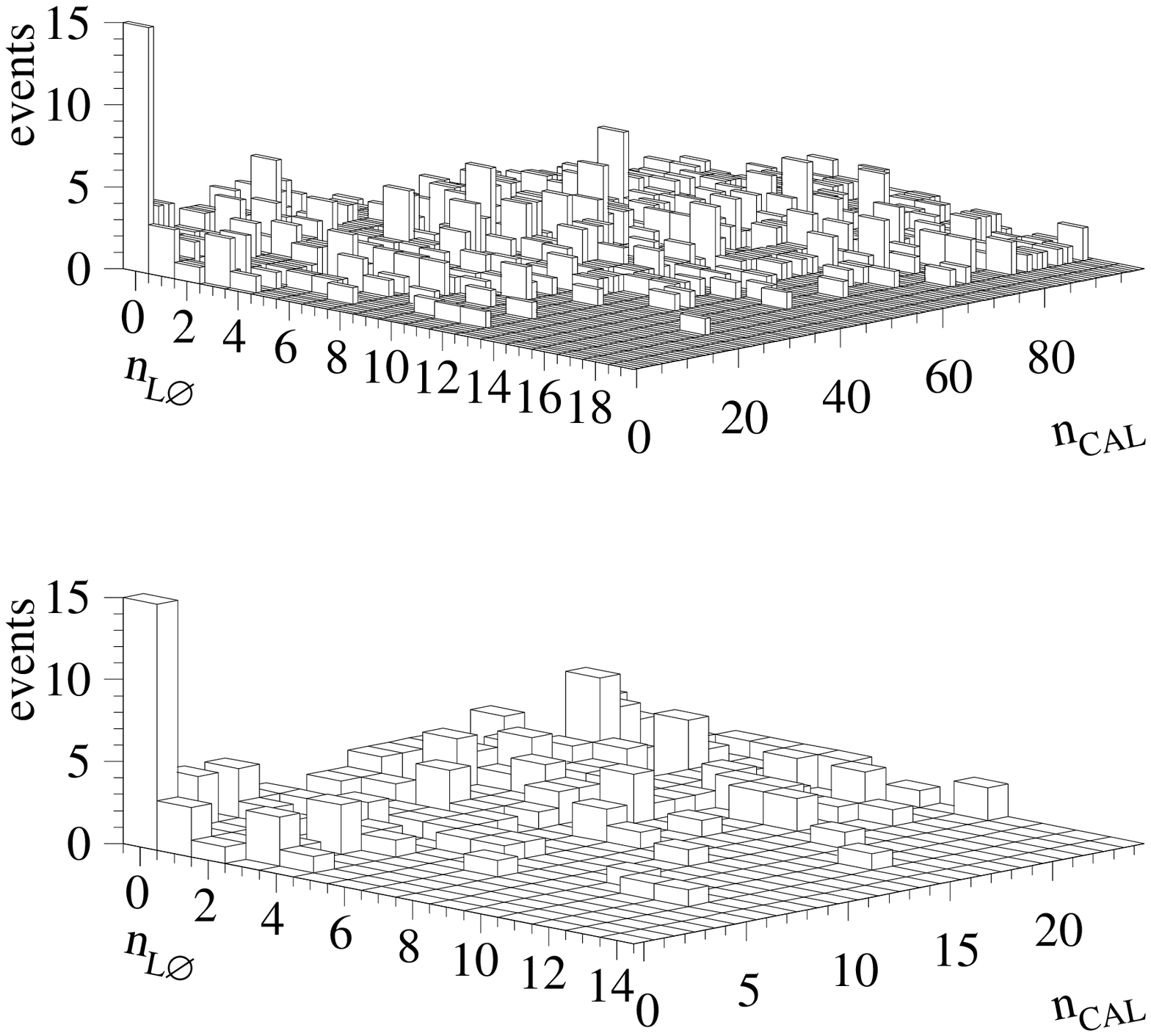}}
\end{center}
\end{minipage}
\hfill
\begin{minipage}{0.45\linewidth}
\begin{center}
\mbox{(b)\epsfxsize 5.5 truecm \epsfbox{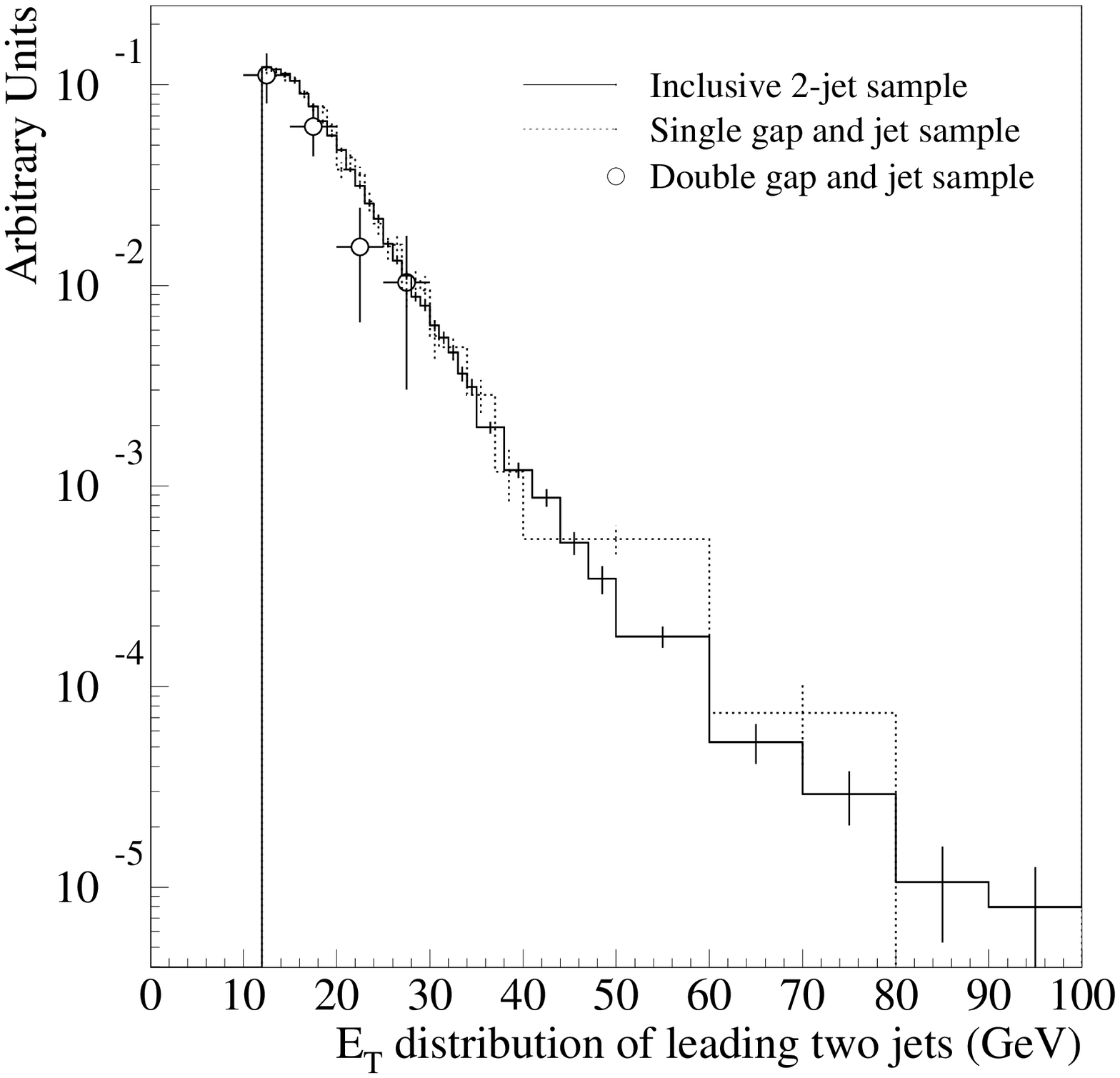}}
\end{center}
\end{minipage}
\caption{(a)
Multiplicity distribution of calorimeter towers ($n_{cal}$) opposite
a tagged rapidity gap for the 630 GeV D\O\  data. The bottom plot shows
an expanded view of the low multiplicity region.
(b) $E_T$ distributions of the leading two jets for three data samples at 630
GeV. An inclusive sample requiring $|\eta_{jet}| < 1.0$ is shown in the
solid histogram. The distribution with the added requirement of one forward 
rapidity gap is shown with dotted lines and the distribution for double
gap events is shown in circles.
} 
\label{fig:d0_dpe_lego_et}      
\end{figure*}   
    
D\O\ has taken inclusive jet data with a special trigger and searched for
dijet events with
two forward rapidity gaps along the direction of the proton and antiproton.
Events were selected having two jets with $E_T > $ 12 GeV, 
$|\eta_{jet}| < 1.0$ and a rapidity gap in the region $2.5 < |\eta| <5.2.$ 
The multiplicity distribution of calorimeter towers and Level{\O}~ hits
($n_{L\O}$), on the opposite side to the rapidity gap, for data taken at 
630 GeV center of mass energy is shown in Figure~\ref{fig:d0_dpe_lego_et} (a).
 A clear peak at low
multiplicity is observed above a fairly flat background in qualitative
agreement with that expected for double pomeron exchange.
Figure~\ref{fig:d0_dpe_lego_et} (b) shows the $E_T$ spectra for the two 
leading jets in
an inclusive sample with two central $E_T > 12$ GeV jets , 
a single forward gap sample and the double gap sample at 630 GeV.
All three spectra are in good agreement where data are available,
implying that the dynamics of leading jets produced in rapidity
gap events appear similar to those of inclusive QCD production.
Similar results are seen by D\O\ in data taken at 1800 GeV.

\section{Conclusions}

Recent studies of hard diffraction at the Tevatron have given new information
about rates of diffraction and dependencies on $E_T$, $\eta$ and \sqs~. 

The fraction of dijet events produced via hard color singlet 
exchange is about 1\% at 1.8 TeV 
and is larger by a factor of 2 to 3 at \sqs = 630 GeV. D\O\ has compared the
$E_T$ and $\Delta\eta$ dependence of the fraction of hard color singlet events
to several models. The data favor a soft-color rearrangement model
preferring initial quark states over 
two-gluon color-singlet models. 

CDF has preliminary results on the momentum distribution of partons in
the pomeron using ``Roman Pot'' detectors to measure quasi-elastic 
scattered $\overline p$ in hard single diffractive events. 
D\O\ has studied hard single diffraction in forward dijet events and 
new results on pomeron parton distributions will soon be available.

Both CDF and D\O\ have preliminary evidence for 
events with a hard double pomeron exchange topology. 
CDF has measured the
fraction of pot dijet gap events to be 0.36\% of the pot dijet events at
\sqs~ = 1.8 TeV. 
D\O\  has studied gap-dijet-gap events
at \sqs = 630 GeV and 1.8 TeV. The $E_T$ distribution of the leading jets
in double pomeron exchange type events is similar to other processes 
producing jets.

More detailed studies of the Tevatron data and further comparisons
with models should give more insight into the nature of hard diffraction.


\begin{references}
\bibitem{geom} The pseudorapidity, $\eta$ is defined as $-\ln{\tan(\theta/2)}$.
  The polar angle, $\theta$ , is defined relative the the proton beam
  direction, $\phi$ is the azimuthal angle, and $E_T$ is the energy
  measured transverse to the beam.
\end{references}
\end{document}